# A Brief survey on Smart Community and Smart Transportation*

Hamid Fekri Azgomi
*IEEE Student member*
ACE Labs
Department of Electrical Engineering
University of Texas at San Antonio
San Antonio, Texas
hamid.fekriazgomi@utsa.edu

Mo Jamshidi
*Fellow, IEEE*
ACE Labs
Department of Electrical Engineering
University of Texas at San Antonio
San Antonio, Texas
mojamshidi4@gmail.com

*Abstract* - **World population growing in conjunction with the preference to live in the cities; make the city management a challenging issue. Traditional Cities with their common features will not be able to handle the human needs. As a result, smart city and its beneficial outcomes will attract a lot of attention recently. In fact, Smart city/community that is the field of collecting and processing data from so many different areas while making proper decisions and feeding to all parts of the system will be the future style of the cities. In other words, smart community is a complex big data problem which should combine different fields of research to make a unit environment. This brief survey will deal with all aspects of the smart communities. There will also be explained different categories of the smart communities in addition to their future challenges. In this work we try to introduce some aspects of the smart transportation in more detail.**

I. INTRODUCTION

Today, more of us than ever before live in cities, on average 60% of the world's population live within 5 kilometer of a city, the future expectation is that this average will rise to 70% by 2030. Based on other report, every week there are more than one million people who are choosing the cities as the place they prefer to live [1]. Primarily, this is due to an enormous growth in urbanization that shows little sign of slowing down. In more developed countries, the number is already far higher, for example in the UK our city living population reflects over 90% of the population [2].

As a primary result of this population growth, there could be seen in lack of efficiency dwindling within cities struggling to manage the growing population. What is more important in this problem there is great capability for accelerating technologies to help carve out a future for everyone on the planet. Improving technologies provide revolutionary change that can usher in quantum-like jumps in civilization. However, the connections between accelerating technologies and metropolitan population growth are complex, certain relationship can be established. For instance, the story from the Industrial Era is that if a machine can replace human workers, then they will [3].

Huge number of people on one hand, and the rapid industrialization in today's world on the other hand raise lots of challenges in the traditional methods of the city management. In other words, it will be impossible to handle all human needs in near future unless we could develop our cities style. The concept of the smart city goes to the combination of the data collection, data analysis, computing, networking and making decision based on this big data flow [4]. One of the key factors that play undeniable role in the smart city is Internet of Things (IoT). There will no smart city without IoT. Based on [5], Internet of Things (IoT) is a system of physical things embedded with sensors, software, electronics and connectivity to provide an infrastructure to perform better by exchanging information with other connected devices, the operator or the manufacturer. Smart cities have become another Information Technology (IT) catchword connected with the IoT, the idea to build a more connected and more collaborative society, an optimized society where data is used to provide stronger insight [5], [6].

From the other perspective, the smart city could be defined as the macroscopic side and the microscopic one. Macroscopic side goes to the decision making which is at the top level of the smart city structure. On the other hand, microscopic analysis is the details of all parts of the smart cities that work together to perform the decision coming from the macroscopic or high-level side. Based on the flow of the data that shared among all parts of the smart cities, this concept is a Big Data system. In the microscopic view point, we could categorize the smart city into different categories. One of the good categorization could be considered as these four: Healthcare , Environment, Energy and Transportation [6]. Each of these four categories consist of different applications that will be mentioned in the following. One of the most important and challenging fields of the smart city is smart transportation. As a prediction, future city management will not be different than the autonomous vehicles in conjunction with the smart traffic control [7], [8]. Autonomous vehicles, smart roads and the infrastructure that all parts of the traffic control could communicate with each other will build the smart transportation in a smart community.

* Partially supported by Air Force Research Laboratory and OSD under agreement number FA8750-15-2-0116.

Recently, there have been lots of research activities that are dealing with different aspects of the smart transportation. Some of these recent trends will be mentioned in this paper.

This paper is structured as follows: in section II; we will present the different technological areas that build the infrastructure of the smart city. Then different categories of smart city will be mentioned in detail. In section III the smart transportation and all its details will be explained. Smart Traffic control management as a challenging issue will be mentioned in this part. In section IV, we present some of open problems in the context of the smart communities that could be considered as the future research directions. Finally, in section V we will provide a brief conclusion of the paper.

## II. SMART CITY AREAS

As we mentioned before, there are various smart city areas we will go through them in this section. We have tried to express the different categories of the smart city fields of research. Some of the current works on each section have been presented in this part.

### A. Healthcare

Recently, with the advantages in the mobile devices and digital tools, there has been possible to establish an environment including connected medical devices under the goal of creating a more comprehensive view of community health and well-being. In this regard, opportunities for physical activity, built environment enhancements, reduced exposure to environmental treats, and expanded availability of affordable, healthful foods could be considered [9].

The use of connected devices and wearable technology has attracted a lot of attention in a world. Here, by allowing people to connect the monitoring of their own health with remotely situated health professionals who can deliver care, the whole continuum of health and wellness is streamlined by increasing efficiency and cost-effectiveness for the patient and the provider.

The intersection between the IoT and block chain technology has emerged as a turning point in the remote patient monitoring field. As a result, it is possible to provide a safe and secure system of vital health data from patients to care providers for analysis and prognosis [10]. So the health care providers could analyze the data they capture to deliver a highly targeted treatment and medication regimen to patients. One sample of the comprehensive healthcare plan in the future smart city could be found in Fig. 1.

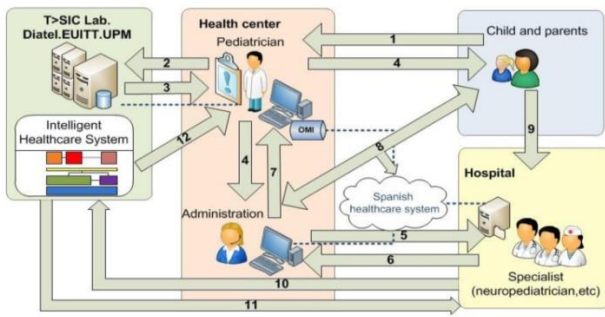

Fig. 1. Smart Healthcare system.

As it could be observed in Fig. 1, there is a real-time connection among all parts of the healthcare system. In this infrastructure, all the activity will be collected by the wearable and mobile devices. In addition, by making a report on one's diet plan, in conjunction with anybody's medical history, there will be a valuable data source for all the citizens. Having access to the hospital and doctor's experiences, we could track all citizens' health status. In such a real-time system, all the diseases could be forecasted and in a timely manner the proper actions will be designed in real time.

In [11] authors are dealing with the structure of the healthcare benefit in the smart city using both mobile and ambient sensors combined with the machine learning. In other words, it emphasizes on the impact of the information and communication technologies (ICTs) to be applied in the healthcare subject. Smartphones and watches come equipped with many sensors. These are some examples of the mobile sensor data gathering devices.

One of the results out came from these data could be used in the design process of the cities [10]. For example, in this paper the urban design based on the neighborhood walkability feature has been introduced. Since the mobile sensor data gathering is not sufficient way, there should be other types of sensors that could collect the data in the situation that one is not convenient with the mobile devices, or in the situations that we need some other types of the data that cannot be collected from just wearable devices. Here is where the effect of ambient sensor technologies. Ambient smart environments are typically a place including sensors that measure environmental features as room temperature, and after making some analysis on them, preparing them for communicating the collected information, and a computer to collect and store the data. One example of these kinds of sensors data collecting could be found in Fig. 2. In this system, some devices ARE attached to the ceiling that monitors heat-based movement with passive infrared (PIR) sensors as well as ambient light levels.

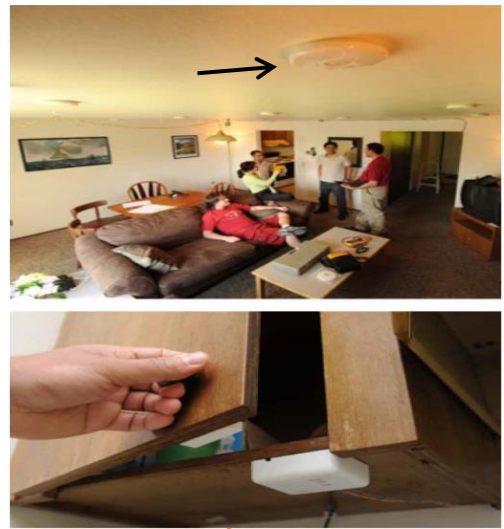

Fig. 2. Common ambient sensors monitor (top) infrared-based motion, ambient light, ambient temperature and (bottom) door open/close status. Arrows indicate a sensor location [11].

And, magnetic sensors are attached to doors and door frames to reflect their open or closure status. In these settings, a wide range of additional sensors can be used [11]. These kinds of environmental sensors could monitor all the activities that could make our data base to be as complete as it is possible.

*B. Environment*

Constructing a city with high environmental standards is highly relied on the use of smart sensor technologies. A key to the recent focus on the future of the public health is the increase in affordability of sensors which could make result in monitoring the city parameters. In past, high-quality sensors were too expensive for most cities that did not make them affordable, but rising demand and advantages in technology have resulted in the availability of more affordable sensors that are able to collect the detailed data needed to make a public health efficiently. In fact, according to a recent report from IDTechEx Research [12], the market for environment sensors is expected to be worth more than $3 billion by 2027. Fig. 3 shows the different aspects of the environmental issues in the future smart city.

Additionally, cities and states are becoming more creative when it comes to financing larger-scale smart city projects such as a network of sensors to measure environmental data. These sensors provide a huge amount of data that can be shared and analyzed to help solve challenges and improve quality of life. According to the World Health Organization, more than 5.5 million people die every year because of air pollution, so increased visibility into pollution levels can create timelier and more accurate warnings that can help save lives [2]. On a smaller scale, data can also be used in more daily city decision-making. For example, data on noise levels might lead to a decision on where to build a school. Or based on the data that came from the air pollution sensors from all parts of the city, it is possible for citizens to change their paths and choose safer ones.

Engineering and material science communities play significant role in keeping the smart cities' environment at a high level of standard by introducing advanced smart materials. Materials with new design and features such as Ultralight weight with high strength aerogel polymers [13] and biocompatible nanocomposites [14] provide the opportunity to produce more advanced and affordable sensors, construction material and biomaterials.

The U.S. Environmental Protection Agency (EPA) is working with cities in Central America, where air pollution levels are extremely high, to use sensors to implement systems -for air quality warnings, which can help people suffering from asthma and other respiratory illnesses adjust their exposure. Some examples are mentioned here.

In Chicago, a network of interactive sensors installed around the city as part of the *Array of Things* project collects real-time environmental data including humidity, cloud cover, noise levels and air quality. The data can be used by the city to develop recommendations that improve sustainability and public health outcomes [5].

Washington, D.C., has been using underground sensors to tackle wastewater and runoff challenges. A new smart wastewater pumping system can change its performance in real time based on conditions detected by the sensors to provide more sustainable outcomes. The sensors also provide feedback to pumping station operators that allows for better, more holistic decision-making and outcomes [12]. In addition, environmental monitoring program, which can be useful in smart cities, are essential to collect important information about the quality of the environment around us [15], [16], [17]. Fig. 4 shows a sample of an air prolusion control system that could collect the data from different sources to prepare it for the upcoming analysis.

Recently, there has been lots of advantaged in street light technologies. Outdoor lighting provides visibility and a safer environment and also they are using lots of energies that make it so important to improve them to some smart fashion. As a result, serving the public with this highly visible infrastructure can be costly. Generally, outdoor lighting consumes 19 percent of energy use. On a local level, up to 40 percent of a city's energy load can be attributed to streetlights [17]. This high value is because of their inefficiency and not having a controlled system for street lights. It is obvious that by making a focus on this area it will cause to have a more clean city with much more financial concerns [9], [17], [18]. Mbarak et al. in [19] have proposed a method for pollution forecasting via machine learning. They have proposed their system to be practical in Northwest Texas. One of their proposed method's advantages is to determine how much green energy should be invested to cut down the pollutants seen on the city [19].

The traditional methods of the street light managing could not solve the future concerns. For example, based on the data that could be gathered from different sources, we could make a data base of the traffic flow in different area and different roads. In detail, let assume that based on the data that comes from the cameras we could make a data base of the traffic flow in different times of the day as well as the days of the year. Based on the prediction algorithms like regression analysis or machine learning approaches, we could predict the flow of the traffic at any time [12].

So, by dimming the light in the days and in the situation that the traffic flow is not too high and increasing the light in the crowded status of the road we could successfully save the energy consumption as well as, enhancing the street light in the desired conditions. One sample of a system including the smart street lights could be observed in Fig. 5.

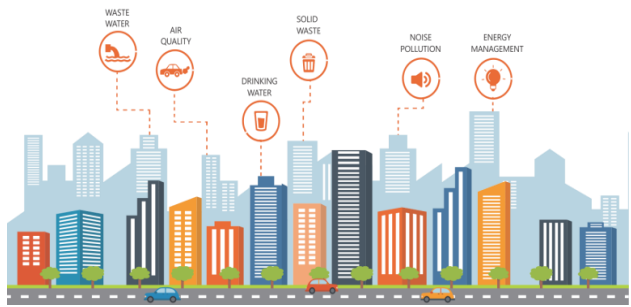

Fig. 3. Environmental Aspects of the Samrt City

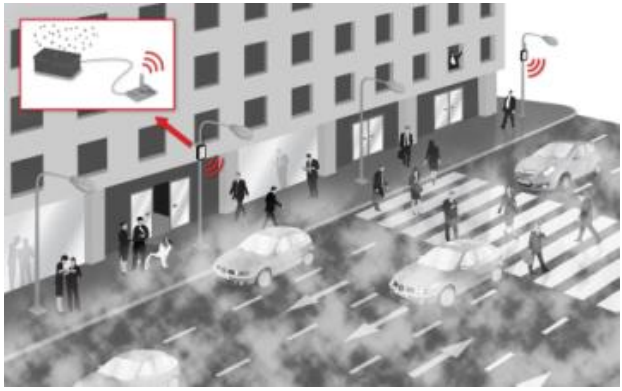

Fig. 4. Air pollution control example

The data that gathered from all sensors in conjunction with the prediction that could be done based on the traffic flow of some specific areas, will go through the control department to make a comprehensive plan for all these smart street lights. As result, big data analysis and cloud-based control system could help the future smart cities have better environmental situation.

Automation is the key to making cities smart. Many cities are used as test beds for experimenting with robots. Robots can be utilized to replace humans or assist them in performing different jobs. Examples include autonomous drones for fast delivering parcels, medical robots for enhancing capabilities of surgeons in hospitals, assistive robots for providing aid for people with limited mobility, and legged robots for performing search and rescue operations [20] [21] [22].

*C. Energy*

A smart city is a sustainable and efficient urban center that provides a high quality of life to its citizens by optimal management of its resources [17]. In the context of the energy in the smart cities, there are two main categories. First, the production of the energy in the future smart cities and secondly, making a comprehensive plan for production, storage and distribution of the energy in a smart community. Calvillo et al. in [17] proposed their work to develop insight in to the complexity of the energy-related activities in a smart community by reviewing advances and trends and then by analyzing the synergies among different intervention areas.

During the recent years, by the advances in lots of research areas, there are high variety of the energy resources that could be produced in different styles. As an illustration, previously the homes in a town were just consumers of the energy; however, in the future world, each home and building has the chance to generate not even its own energy, but also it can have the plan to sell the produced energy. Having a plan and a schedule for all parts of the smart city could significantly cause a huge improvement in the context of the energy in the smart communities. The data that should be collected from all parts of the smart city are including the production capacity, storage capacity and a plan that determines how each source could distribute energy among all parts of the smart city. From the data side, it is big data problem that needs to consider so many parameters. One example of such a huge system that based on the collected data from all parts of smart city could be found in Fig. 6.

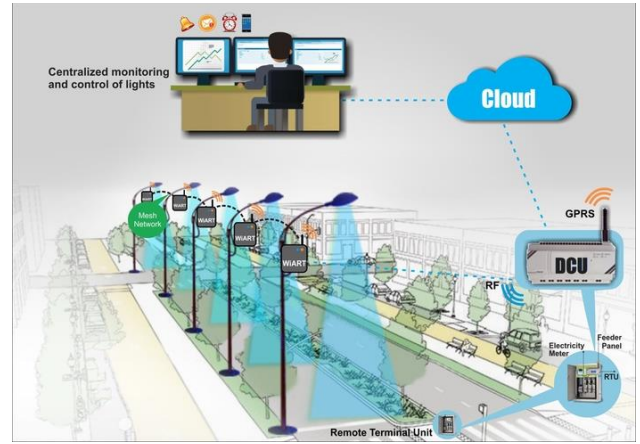

Fig. 5. Smart street light

As we can see in Fig. 6. there are so many parts in the future smart city that can talk to each other; from autonomous vehicles and their required energy usage to the generators, from the renewable energy producers to the smart consumer/producer homes.

One of the main research challenges related to Distributed Generation (DG) is determining of the optimal configuration, location, type and sizing of the generation units, so that the system provides all energy requirements at an optimized cost manner [23]. Upadhyay et al. in [24] and Chauhan et al. in [25] have reviewed the recent activities on DG programming and control plan.

One of the most infrastructures of the smart grids as a main part of the smart cities is sensing concept [26]. This kind of a sensor is the one that is much more advanced and has great features. As an illustration, these types of advanced sensors not only make the required measurements, but also, they help to analyze the data that gathered collected and prepare them for the further analysis. Such sensing facilities will have the task to assure a safe and efficient delivery of electricity even when there is the extra energy on demand. Real-time information concerning the local energy consumption/production and additional information on the power quality of energy transiting in each node of the electric network will allow the smart power grid to manage promptly unpredictable events such as a sudden increase of energy demand or the sudden failure of distributed generation or even sudden changes in the climate changed which will affect the energy production level in different ways.

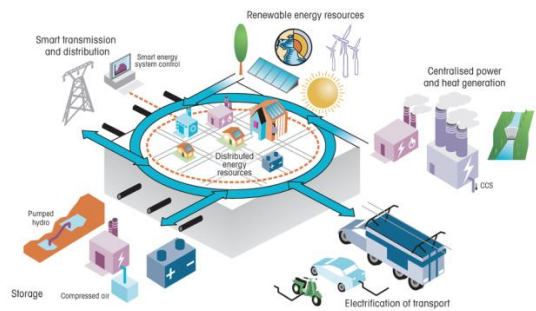

Fig. 6. Smart Energy management system

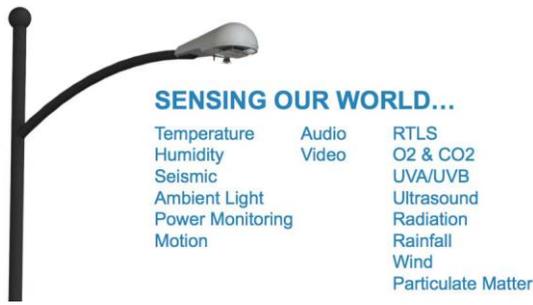

Fig. 7. Different sensors for a street lighting in a smart city [18].

Morello et al. in [22] have dealt with this new generation of the sensors and introduce some of their application in the various parts of the smart cities including street lighting, automotive and smart transportation. Fig. 7 shows one of these kinds of advanced sensors which could be used in the smart street lights.

As it can be observed in Fig. 7. this kind of sensors not even helps in the construction of our comprehensive smart grid, but also can have lots of advantages in different aspects of the smart community decision making side.

*D. Trasportation*

The most important and challenging part of the smart city is anything that related to the smart transportation. It is obvious that with the population growing and their needs for more advance transportation facilities; on one hand, and the energy consumption on the other hand will lead the future world to establish much more powerful transportation systems.

In other words, most notable problems are [27]:

- Traffic congestion
- Environmental impacts
- Energy consumption
- Accidents and safety
- High maintenance cost
- Land consumption

To address these issues, the concept of smart transportation or intelligent transportation system has been presented [8]. In fact, an intelligent transportation system (ITS) is an advanced application which without embodying intelligence aims to provide innovative services relating to different modes of transport and traffic management and enable users to be better informed and make safer, more coordinated, and 'smarter' use of transport networks.

The most important issues in smart transportation could be considered as time and energy. In the Time concern, it is supposed to save the citizens' time while they are in their trips. This means that we should have a structure to optimize the arrival time in every day travels. More specifically, we should improve the traditional traffic signs (including the traffic lights) to save the time that is passed in citizens daily schedules. On the other side, based on a research near 30% of the US energy usage belongs to the transportation [28]. It means that working this area will significantly reduce total energy usage in the future smart cities.

The topics that could be considered in this area are autonomous vehicles, electric vehicles and smart traffic control management. Based on the last part of the introduction, smart transportation including the autonomous vehicles and smart traffic control system will be explained in detail in the following session.

### III. SMART TRANSPORATION

Smart transportation in smart community consists of two main parts; smart traffic control management and autonomous vehicles. The ultimate goal of the smart traffic control is to have a system in which all the vehicles and traffic signs and control bases could share the data among themselves to make a proper decision in a secured and optimized environment. In this regard, we should make a deep attention to the concept of the electric vehicles. Future transportation will be highly relying on the driver-less or autonomous vehicles which do not need the humans to make decisions for them. In a cloud-based system all parts of the smart transportation systems will share all required data among themselves and then by doing appropriate analysis, best decisions would be made. Some structure of the smart traffic control using the autonomous vehicles which relate to each other could be found in Fig. 8.

As it can be seen in this figure, all the cars, traffic control facilities as the traffic signs and traffic light are in a cloud-based connection among them. They can send and receive data to construct a powerful data base. There are lots of details in this regard that are active research areas in the field of smart traffic control of the autonomous vehicles. Image detection, traffic flow prediction and defining different optimization problem are some of them [29].

One of the other important requirements in the smart transportation system is the smart traffic signs and smart roads. In a smart city all the traffic signs and the road facilities can make decisions based on the data came from the vehicles and the cameras that cover all the roads and intersections. In a real-time manner, all the data from the start point, destination point, and the traffic flow will be collected and based on the computations that are supposed to be done on them; the proper decisions will be sent to all the elements in a smart transportation system. One example of a smart traffic control management system could be observed in the Fig. 9. In what follows we will review some of the trends that have been done in this regard.

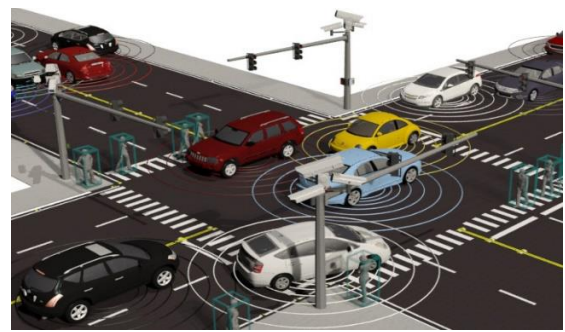

Fig. 8. Smart transportation scheme.

Shahidehpour et al. in [28] have proposed a bi-level optimization framework to settle the optimal traffic signal setting problem. The upper-level problem determines the traffic signal settings to optimize the overall trip time; on the other hand, the lower level problem is for achieving the network equilibrium using the settings calculated at the upper level. The generic algorithm has been used to decouple the upper level and the lower level in the complex problem. Younis et al. in [30] have worked on the control of the traffic light in the intersections. It is considered as a cyber-physical system consists of the lights management and the vehicles. They have proposed a novel approach to achieve a fixed periodic setting for all the lights that could be categorized to different times of the day (Morning, Noon and Night). The approach that has been proposed in this paper is based on the data that collected from all the intersections. Their main focus is on the reducing the time that the vehicles wait in the traffic flow. Estrada et al. in [26] presented their method in order to reduce the cognition in the traffic flow. They have proposed a multi-objective system to optimize the trip time by means of a proper design of the duration of the green traffic light in the intersections.

One of the important areas in the smart traffic control field of research is related to the ability of the prediction of the control flow based on the data that are collected from the limited number of the cameras. Beside the limited number of cameras that should be installed in the intersections, there is other issue with the camera quality. Katsuki et al. in [31] have proposed a method to estimate the traffic flow based on the limited number of low-quality cameras. Their approach features two main algorithms. The first is a probabilistic vehicle counting algorithm from low-quality images that falls into the category of unsupervised learning. The other is a network inference algorithm based on an inverse Markov chain formulation that infers the traffic at arbitrary links from a limited number of observations. The main advantage of their proposed approach compared to the traditional methods is the cost benefits. Their machine learning approach leads them to do it in a reasonable cost. Duan et al. in [32] dealt with traffic flow prediction problem based on deep architecture models with big traffic data. In this paper, a deep-learning-based traffic flow prediction method is explained, which considers the spatial and temporal correlations inherently.

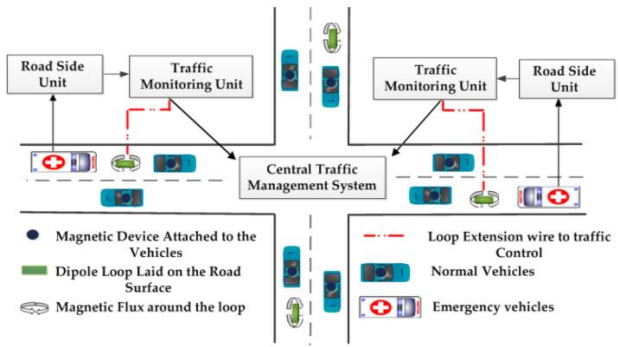

Fig. 9. Smart traffic control management.

A stacked auto encoder model is used to learn generic traffic flow features, and it is trained in a greedy layer wise fashion. Their proposed method can successfully discover the latent traffic flow feature representation, such as the nonlinear spatial and temporal correlations from the traffic data.

Arezou et al. in [33] they have proposed a machine learning approach to predict the traffic flow both in offline and online mode. They have proposed two different algorithms; Multi-Layer Perception (MLP) that is trained using the yearly data for offline prediction and stochastic gradient descent which is employed in online forecasting.

In the concept of the smart traffic control, one of the challenging issues that attract lots of attention is path planning analysis [34]. Path planning in a large environment which consists of autonomous vehicles as smart traffic signs is a complex and big data problem. As we are dealing with a huge data problem, the traditional approaches will not be helpful [35], [36]. The main reason is the nature of the problem which is based on a real-time connection among all parts of the system. To illustrate this issue, we are going to review this problem in what follows.

Nambiar et al. in [37] have proposed the game theory approaches in solving the multi agents path planning problem. In their work, they have also compared it with the traditional A* approach. Authors in [30] proposed the A* algorithm in solving the path planning problem considering the stochastic obstacle avoidance. Their focus is on single agent path planning problem.

In [38] Rodrigues et al. have proposed a method that is related to the behavior selection for autonomous vehicle for solving the speed planning. Their method results learning human driver's longitudinal behaviors for driving at a non-signalized roundabout. This knowledge is then used to generate longitudinal behavior candidate profiles that give the autonomous vehicle different behavior choices in a dynamic environment. A decision making algorithm is then employed to tactically select the optimal behavior candidate based on the existing scenario dynamics.

Consider the situation that the driver-less cars are going through the roads and intersections, based on the traffic flow, all data collected from the cars and their destinations and the traffic constraints, the proper path planning should be set and based on that, the appropriate traffic control systems will be organized. In the future smart traffic control, the ultimate goal is to remove the traffic signs performance and minimize the traffic concerns (such as the time and distance). The traffic optimization could be considered as removing the red light, smart stop signs, and smart speed limit and so on. Based on what were mentioned in this section, our future challenge will be formulating the path planning problem in a real-time environment and trying to solve it in an advanced fashion that combine the path planning problem in conjunction with the smart roads and smart traffic control systems.

IV. OPEN PROBLEMS

In this section we try to address some of the active challenges in the smart community field of research. So, we have proposed some of future smart city challenges with respect to the data perspective. Based on the categorization that we had before, we will mention them in the four areas.

## A. Healthcare

Future healthcare system will be highly around the data analysis. AI techniques and data modelling will play undeniable roles in this regard. In each patient categories, their daily activities, exercise habits, diets and comparing them with the symptoms of each disease at different stages, will have a significant achievement in the smart healthcare systems. Base on the huge amounts of the parameters in this problem, there may be beneficial to apply machine learning approaches to have better vision and modelling. Designing a cloud-based system among the patients, doctors/nurses and the hospitals will attract lots of attention in the future smart communities. Working on a unified platform among all parts of the future healthcare system could be challenging.

Another challenging research topic could be done on the microscopic point of view in the healthcare system. Future medical systems will be developed around the data analysis. Collecting data from all diseases and trying to fit some advance model for them would be challenging. Based on the high verity of the parameters in medical studies, these kinds of model are highly complex. As a result, machine learning approaches could be beneficial here.

## B. Environment

Based on the advances on sensor technologies, there could be interesting and important challenge to work on sensory data fusion on all aspects of the smart environment issues. As there will be some limitation on the number of sensors that would be used, some hot topics could be finding some predication based on the data that came from the limited number of sensors. Some of the applications could be found in what follows.

In smart community's environmental issues, there could be some study on the smart water purifying methods which it means by collecting the data from whole duration of the year, knowing the water ingredients at each location and defining some purifying algorithm to manage the water resources efficiently. Another issue could be considered as the smart water distribution plan which could be applied in the farm irrigation. Again, the main requirement is to monitor the precipitation and try to predict the pattern and then make the proper plan for farm irrigation. AI techniques including machine learning approaches could be helpful in this regard.

Another challenging issue could be the air pollution problem. There is no doubt that in the future structures of the huge cities, there could be much more efficient systems taking care of the air purifying. Finding some patterns based on the city activities that may cause the pollution could be beneficial in designing the proper system. By monitoring and analyzing the related data, it would be possible to predict the upcoming amounts of the pollutant in detail. Having knowledge about the detail of the air pollution will lead us to design more professional purification structures.

In addition, in the contexts of the smart street lights, some AI techniques could be useful to extract the data from all city streets. There are lots of different factors that are important in having a comprehensive data base-i.e., Weather condition, Time of the day, Time of the year and so on [19].

Energy

From the data perspective, as it was mentioned before, the world will be around the green energy resources. For example, based on some recent U.S. government rules, in California, all the houses are required to install the solar cells. In this regard, having a comprehensive pattern that tell us about the upcoming produced energy for every units, the energy usage, storage capacity and distribution capabilities will undoubtedly be critical. Because of the high number of parameters that are important in this problem, it should be studied as a complex one which smart approaches could be applied significantly. Like the smart healthcare system, working on a unified platform among all parts of the future smart grids could attract lots of attention in near future.

## C. Transportation

In this area there could be lots of challenges that could be considered. In what follows, some of them may be found,

✓ Multi-sensory data fusion for the autonomous vehicles,

✓ Traffic flow prediction,

✓ Cloud-based multi agents path planning,

✓ Real-time path planning problem in a smart community considering the interaction among all smart city agents

On the top of all smart community open problems, one challenging issue that will attract lots of attention in the following years, is the concept of the security. Since the smart city infrastructure is IoT, cyber-attacks will be a vulnerable threat. So, providing the secured environment could be one of the other future important challenges in practical smart community implementation.

## V. CONCLUSION

In this survey paper, an attempt made to go through the concept of smart city and smart community. Its different aspects were reviewed and some of their challenges presented. In addition, it indicated that among all categories including Environment, Healthcare, Energy and smart transportation, the latter one has received more attention in recent years. Some recent work on smart traffic control and autonomous vehicles were presented and finally complex path planning framework was discussed as our future challenge for smart cities. At the end, we tried to introduce some of the future challenges in this concept.

### Acknowledgement

This paper is supported, in part, by Air Force Research Laboratories and OSD for sponsoring this research under agreement number FA8750-15-2-0114.